\documentclass[twocolumn]{autart}    

\usepackage{graphicx}          
                               
\usepackage{cite}
\usepackage{epsfig}
\usepackage{float}
\usepackage{subfigure}
\usepackage{times}
\usepackage{amsmath}
\usepackage{amssymb}
\usepackage{multirow}
\usepackage{stfloats}
\usepackage{color}
\usepackage{cite}
\usepackage{epsfig}
\usepackage{float}
\usepackage{subfigure}
\usepackage{times}
\usepackage{amsmath}
\newtheorem{theorem}{Theorem}
\newtheorem{lemma}{Lemma}
\newtheorem{remark}{Remark}
\newtheorem{definition}{Definition}
\newtheorem{assumption}{Assumption}
\usepackage{amssymb}
\usepackage{multirow}
\usepackage{hhline}
\usepackage{arydshln}
\begin{document}

\begin{frontmatter}

\title{Distributed Multi-Time Slot Power Balancing Control of Power Systems with Energy Storage Devices \thanksref{footnoteinfo}} 

\thanks[footnoteinfo]{This paper was not presented at any IFAC 
meeting. Corresponding author T. Liu. This work was supported by the Research Grants Council of the Hong Kong Special Administrative Region under the General Research Fund Through Project No. 17209219.}

\author[hku]{Luwei Yang}\ead{lwyang0804@outlook.com},    
\author[hku]{Tao Liu}\ead{taoliu@eee.hku.hk},               
\author[hku,Rome]{David J. Hill}\ead{davidj.hill@monash.edu}  

\address[hku]{Department of Electrical and Electronic Engineering, University of Hong Kong, Hong Kong}  
\address[Rome]{Department of Electrical and Computer Systems Engineering, Monash University, Melbourne, 3800, Australia}             


\begin{abstract}                          
This paper studies a crucial problem in power system balancing control, i.e., the multi-time slot economic dispatch (MTSED) problem, for power grids with substantial renewables, synchronous generators (SGs), and energy storage devices (ESDs). The target of MTSED is to optimally coordinate active/reactive power outputs of all controllable units to meet a forecast net demand profile over multiple time slots within a receding finite time horizon. Firstly,  the MTSED is formulated as an optimization problem with operational constraints, including the limits on the output of each controllable unit, ramping rates of SGss, energy levels of ESDs, and bus voltages.  Then, a novel projection-based algorithm is developed to solve the problem in a distributed way.  In particular, the distributed algorithm is not limited to solving the MTSED problem but also applies to more general optimization problems with both generic convex objective functions and hard feasibility constraints.  Finally,  case studies verify the effectiveness of the proposed method.
\end{abstract}

\end{frontmatter}
\section{Introduction}
To maintain the real-time power balance between generation and demand is critical for operating a stable electric power system \cite{15}. This key task is practically achieved by implementing some sophisticated power balancing control approaches such as unit commitment (UC), economic dispatch (ED), and load frequency control (LFC), which operate on different time scales \cite{oc1}. Among these methods, ED has received considerable attention. It is to keep the power balance by optimally tuning power set-points of synchronous generators (SGs) for each certain time period (e.g., $10$ or $15$ minutes) and plays a crucial role in power system frequency regulation. However, for a modern power system with high penetration of renewables, ED is facing new challenges caused by the fact that SGs with limited ramping capabilities may be inadequate to follow the fast fluctuations of renewables \cite{re}. 

To cope with these challenges, energy storage devices (ESDs) have been advocated to participate in ED owing to their instantaneous responsiveness and low emissions \cite{es1}. Nevertheless, these devices have limited energy capacities and need to closely cooperate with SGs. This presents an urgent need to develop effective ED strategies for systems with both SGs and ESDs by fully considering their respective properties. In addition, the conventional ED problem considers power dispatch of SGs only in a single time slot \cite{15} and may not apply to systems with ESDs directly \cite{xj}. This is because the operating conditions of these energy-constrained units are highly coupled with time by the charging/discharging dynamics. 

To address this issue, the multi-time slot ED (MTSED) problem is introduced. It studies the optimal power dispatch of SGs and ESDs across multiple time slots within a receding finite time horizon and hence has the ability to handle time-coupling constraints of the system (e.g., the ramping limits of SGs and energy capacity limits of ESDs) \cite{dis4}. Moreover, MTSED determines the power allocation of all controllable units by taking advantage of both the accuracy of short-term forecasts of renewable generation and load demand as well as their trends from the long-term predictions. As a result, it may achieve a more economical and secure balancing control than that of the single-time slot ED. 

Due to its advantages, the MTSED problem has attracted increasing attention recently. Numerous centralized algorithms have been developed to solve MTSED in the literature (see \cite{medr} and references therein). A centralized model predictive control-based algorithm and Lagrangian relaxation~based~algorithm are proposed in \cite{ca1,ca2}, respectively. A centralized dual decomposition algorithm is designed to solve the stochastic MTSED problem that considers forecast uncertainties of renewable generation in \cite{ca3}. However, these centralized approaches may cause a heavy communication burden for the control centre and are vulnerable to single-point failures \cite{cen}.

It has been pointed out in \cite{copf} that the MTSED with ESDs can also be solved by distributed algorithms which guarantee asymptotic convergence of the same optimal solutions as their centralized counterparts (see \cite{yipeng} for a detailed review on distributed optimization algorithms). Some related results by using distributed algorithms to solve the MTSED problem have been reported in the literature (e.g., \cite{dis1,dis2,dis3,dis4,es1,yiguang}). In \cite{dis1,dis2}, the ESDs are treated as traditional SGs with upper and lower power bounds, but the energy capacity limits are not considered. References \cite{dis4,dis3} take the energy constraint into account but assume the unity storage charging and discharging efficiencies which are usually different in practice. In addition, the physical constraints on bus voltages and reactive power injections are not considered in the abovementioned works, which may make the obtained results infeasible for practical implementations \cite{chao}.

Further, the existing studies \cite{dis1,dis2,dis3,dis4,yiguang} require the strict (or strong) convexity assumption of the objective functions, but sometimes the objective function is only convex. For example, a linear cost function is commonly used for ESDs in power system control and analysis  (e.g., \cite{ess1,ess2,ess3}). However,  designing a distributed solution algorithm for problems with generic convex objective functions is still challenging \cite{yiguang}.

In view of the abovementioned issues, this paper studies the  MTSED of power systems with both SGs and ESDs. The MTSED task is formulated as a constrained optimization problem with a generic convex storage cost function and multiple system operational constraints, including bus voltage, power/ramping capacity of SGs, and power/energy capacity of ESDs. The decoupled linearized power flow (DLPF) model proposed in \cite{lpf} is employed to introduce bus voltages and reactive power injections into the studied problem.  Further, a novel distributed projection-based algorithm is developed to solve the MTSED problem by fully considering the different characteristics of each controllable unit.  Also, theoretical proofs of asymptotic convergence of the designed algorithm are provided. 

The rest of this paper proceeds as follows. Firstly,  Section \uppercase\expandafter{\romannumeral2} formulates the MTSED problem.  Then,  Section \uppercase\expandafter{\romannumeral3} gives the distributed projection-based algorithm for MTSED with some necessary preliminaries, where theoretical proofs of optimality and convergence of the proposed algorithm are provided.  Case studies are conducted in Section \uppercase\expandafter{\romannumeral4} using a modified IEEE 14-bus test system. Finally, conclusions are drawn in Section \uppercase\expandafter{\romannumeral5}.

\textit{Notations:} Denote the sets of real numbers, $n$-dimensional real vectors, $(m\times n)$-dimensional real matrices by $\mathbb{R}$, $\mathbb{R}^n$~and $\mathbb{R}^{m\times n}$, respectively. Denote the sets of non-negative integers, non-negative real numbers, and $n$-dimensional vectors with all non-negative real entries by $\mathbb{Z}_+$, $\mathbb{R}_+$ and $\mathbb{R}^n_+$, respectively. Let $I_n$ be the $n$-dimensional identity matrix; $0_{m\times n}$ be the $(m\times n)$-dimensional zero matrix; and $1_n$ be the $n$-dimensional vector with all entries equal to $1$. We use the notations $\text{diag}(a_1,\dots,$ $a_k)$ and $\text{diag}(A_1,\dots, A_k)$ to represent the diagonal and block diagonal matrices with diagonal entries being $a_i\in\mathbb{R}$ and $A_i$ $\in\mathbb{R}^{m_i\times n_i}$, $i=1,\dots,k$, respectively. Moreover, the notation $\Vert \xi\Vert$ denotes the Euclidean norm of vector $\xi\in\mathbb{R}^n$, and $A_1\otimes A_2$ is the Kronecker product of matrices $A_i\in\mathbb{R}^{m_i\times n_i}$, $i=1,2$. We denote $S_1\times\dots\times S_k$ as the Cartesian product of sets $S_i\subset\mathbb{R}^{n_i}$, $i=1,\dots,k$, and $\text{col}(\xi_1,\dots,\xi_k)=(\xi^T_1,\dots,\xi_k^T)^T$ as the column vector with $\xi_i\in\mathbb{R}^{n_i}$, $i=1,\dots,k$. For vectors $\xi=\text{col}(\xi_1,\dots,\xi_n)$, $\eta=\text{col}(\eta_1,\dots,\eta_n)\in\mathbb{R}^n$, we denote $\xi\leq \eta$ if $\xi_i\leq\eta_i$ for all $i=1,\dots,n$.

\section{Problem Formulation}

Consider a connected power transmission network with $n$ buses where each bus $i\in\mathcal{N}$ with $\mathcal{N}=\{1,\dots,n\}$ can~be equipped with one or more devices such as a renewable generating unit, SGs, ESDs, and non-dispatchable load, or can be just a connection bus without any device. We denote $\mathcal{N}_g$ and $\mathcal{N}_s$ as the index sets of buses with SGs and ESDs, respectively, and allow the situation that a bus $i$ can have both SGs and ESDs, i.e., $i\in\mathcal{N}_g\cap\mathcal{N}_s$. We assume the renewable generation outputs and load demand profiles at buses can be predicted over a  receding finite time horizon (or a prediction window), which is divided into $\tau$ time slots with identical duration $T_o$, i.e., $t\in[hT_o,(h+\tau)T_o)$, $h\in\mathbb{Z}_+$. We further denote $\mathcal{T}=\{1,\dots,\tau\}$ as the index set of the time slots in each prediction window.

Within a given prediction window starting at $t=hT_o$, for each bus $i\in\mathcal{N}$ and time slot $k\in\mathcal{T}$, let $p_{gi}[k]$, $q_{gi}[k]$ be the active and reactive power outputs of SGs; $p_{ci}[k]$, $p_{di}[k]$ be the charging and discharging powers of the ESD; $d_{pi}[k]$, $d_{qi}[k]$ be the active and reactive power components of the net load demand (i.e.,~the~non-dispatchable load minus renewable generation); $v_{i}[k]$, $\theta_{i}[k]$ be the bus voltage magnitude and phase angle, respectively. For a bus that has no SG or ESD, we can simply set $p_{gi}[k]=q_{gi}[k]=0$, $\forall i\in\mathcal{N}\setminus\mathcal{N}_g$, and $p_{ci}[k]=$ $p_{di}[k]=0$, $\forall i\in\mathcal{N}\setminus\mathcal{N}_s$.  We assume that all these variables keep unchanged within each time slot $k$.

The control target of MTSED is to optimally allocate~the active/reactive powers $p_{gi}[k]$, $q_{gi}[k]$ of each SG and charging/discharging powers $p_{ci}[k]$, $p_{di}[k]$ of each ESD such that the system net demand can be supplied at all time slots in each prediction window with the operational constraints of the system satisfied.  Similar to MPC, the MTSED problem is solved at each $t=hT_o$ over the prediction window $t\in[hT_o,(h+\tau)T_o)$, but only the solution for the first time-slot, i.e., $p_{gi}(hT_o)=p_{gi}[1]$, $q_{gi}(hT_o)=q_{gi}[1]$, $i\in\mathcal{N}_g$ and $p_{ci}(hT_o)=p_{ci}[1]$, $p_{di}(hT_o)=p_{di}[1]$, $i\in\mathcal{N}_s$, is applied to the controllable units connected to bus $i$ in the current time slot $t\in[hT_o,(h+1)T_o)$. This process is repeated for the next time step $t=(h+1)T_o$ with the latest forecasts of renewable generation and load demand.  More precisely, at each $t=hT_o$, the MTSED problem is formulated as follows
\begin{subequations}\label{op}
\begin{align}
&\hspace{-0.6cm}\mathop{\text{min}}_{\mathop{}_{p_{gi}[k],q_{gi}[k],p_{ci}[k],p_{di}[k]}}\sum_{i=1}^n\sum_{k=1}^{\tau}(f_{gi}(p_{gi}[k])+f_{si}(p_{ci}[k],p_{di}[k]))\nonumber\\
\text{s.t.}~&p_{gi}[k]-d_{pi}[k]-p_{ci}[k]+p_{di}[k]\nonumber\\
&=\sum_{j=1}^n (g_{ij}v_{j}[k]-b_{ij}'\theta_{j}[k])\label{c1}\\
&q_{gi}[k]-d_{qi}[k]=-\sum_{j=1}^n(b_{ij}v_{j}[k]+g_{ij}\theta_{j}[k])\label{c2}\\
&\underline p_{gi}\leq p_{gi}[k]\leq\overline p_{gi}\label{c3}\\
&\underline q_{gi}\leq q_{gi}[k]\leq \overline q_{gi}\label{c4}\\
&\underline r_{i}T_o\leq p_{gi}[k]-p_{gi}[k-1]\leq \overline r_{i}T_o\label{c5}\\
&0\leq p_{ci}[k]\leq \overline p_{ci}\label{c6}\\
&0\leq p_{di}[k]\leq \overline p_{di}\label{c7}\\
&\underline c_{i}\leq c_{i}[0]+T_o\sum_{l=1}^{k}(\eta_{ci} p_{ci}[l]-\eta_{di}^{-1} p_{di}[l])\leq\overline c_{i}\label{c9}\\
&\underline v_{i}\leq v_{i}[k]\leq \overline v_{i},\label{c10}
\end{align}
\end{subequations}
for all $i\in\mathcal{N}$, $k\in\mathcal{T}$, where $p_{gi}[k]$, $q_{gi}[k]$, $p_{ci}[k]$, $p_{di}[k]$ are the decision variables. 

For the SG at bus $i$, $i\in\mathcal{N}_g$, $f_{gi}(p_{gi}[k])$ is the generation cost at time slot $k$; $\underline p_{gi}$, $\overline p_{gi}$ ($\underline q_{gi}$, $\overline q_{gi}$) are the minimum and maximum active (reactive) power outputs; $\underline r_{i}$, $\overline r_{i}$ are the ramp-down and ramp-up limits, respectively.  For the ESD at bus $i$, $i\in\mathcal{N}_s$, $f_{si}(p_{ci}[k],p_{di}[k])$ is the operation cost at time slot $k$; $\overline p_{ci}$, $\overline p_{di}$ are the maximum allowable charging and discharging powers; $\eta_{ci}$, $\eta_{di}$ are the charging and discharging efficiencies that are usually different from each other; $\underline c_{i}$, $\overline c_{i}$ are the lower and upper bounds on the stored energy, respectively.  For each bus $i$, $i\in\mathcal{N}$, $\underline{v}_i$, $\overline{v}_i$ are the amplitude constraints of the bus voltage; coefficients $g_{ij}$, $b_{ij}$, $b_{ij}'$ are the conductance, susceptance, and susceptance without shunt elements between bus $i$ and bus $j$, $j\in\mathcal{N}$,~respectively.  In what follows, we define $G=[g_{ij}]$, $B=[b_{ij}]$, $B'=[b_{ij}']\in$ $\mathbb{R}^{n\times n}$ as the conductance matrix, susceptance~matrix~and~susceptance matrix without shunt elements of the entire grid.  For details of the matrices $G$, $B$, and $B'$ of a transmission network, please refer to \cite{lpf}.

In the constraints \eqref{c5} and \eqref{c9}, when $k=1$, $p_{gi}[k-1]=p_{gi}[0]$ and $c_{i}[k-1]=c_{i}[0]$ denote the active power generation and energy level of the corresponding devices right before $t=$ $hT_o$. Furthermore, for those buses $i\in\mathcal{N}\setminus\mathcal{N}_g$ that  have no SG, we set $\underline{p}_{gi}=\overline{p}_{gi}=\underline{r}_i=\overline{r}_i=0$; and  for buses $i\in\mathcal{N}\setminus\mathcal{N}_s$ that have no ESD, we set $\eta_{ci}=$ $\eta_{di}^{-1}=\overline{p}_{ci}=\overline{p}_{di}=\underline{c}_i=$ $\overline{c}_i=0$.

The objective function is to minimize the total operation cost of all SGs and ESDs, in each prediction window. Constraints \eqref{c1}, \eqref{c2} represent the active and reactive power balance constraints at each bus. Constraints \eqref{c3}, \eqref{c4} describe the active and reactive power limits of SGs, respectively, and \eqref{c5} denotes the SG ramping limit. The ESD charging and discharging powers are respectively bounded by \eqref{c6} and \eqref{c7}, and the energy levels of ESDs are bounded by \eqref{c9}.  Finally, constraint \eqref{c10} is to make the bus voltage magnitude stay within its own admissible range.

It should be noted that an ESD cannot be charged and discharged simultaneously. This can be ensured by introducing the nonlinear constraint $p_{ci}[k]p_{di}[k]=0$, which will increase difficulties for solving problem \eqref{op}. We will show later that by choosing a proper cost function for ESDs,  this physical constraint can certainly be satisfied (see Remark \ref{remark1} for details). As a consequence, we do not include the constraint $p_{ci}[k]p_{di}[k]=0$ in problem \eqref{op} and leave this issue to the selection of the cost function for ESDs.

For the objective function, we adopt the following standard quadratic cost function for each SGs \cite{oc1} 
\begin{equation}\label{gc}
f_{gi}(p_{gi}[k])=\frac{a_{gi}}{2}  p^2_{gi}[k]+b_{gi} p_{gi}[k]+c_{gi}
\end{equation}
where $a_{gi}\in\mathbb{R}_+$, $b_{gi},c_{gi}\in\mathbb{R}$, $\forall i\in\mathcal{N}_g$; and $a_{gi}=b_{gi}=c_{gi}=0$, $\forall i\in\mathcal{N}\setminus\mathcal{N}_g$. In addition, we use the following linear operation cost function of both charging and discharging powers for each ESD, which is extensively adopted in energy storage systems (e.g.,  \cite{ess1,ess2,ess3})
\begin{equation}\label{esc}
f_{s_i}(p_{ci}[k],p_{di}[k])=a_{si}(p_{ci}[k]+p_{di}[k])+b_{si}
\end{equation}
where $a_{si}\in\mathbb{R}_+$, $b_{si}\in\mathbb{R}$, $\forall i\in\mathcal{N}_s$; and $a_{si}=b_{si}=0$, $\forall i\in\mathcal{N}\setminus\mathcal{N}_s$. For details of the~physical~meanings of the generation and storage cost functions \eqref{gc} and \eqref{esc}, please refer to \cite{oc1,ess3}, respectively.

\begin{remark} \label{remark1} It has been shown in \cite{ess2} that the function \eqref{esc} is monotonically increasing, and hence can prevent the charging power $p_{ci}[k]$ and discharging power $p_{di}[k]$, $i\in\mathcal{N}_s$, from being simultaneously non-zero, which guarantees that ESDs can only operate either in the charging or discharging mode at any time.  
\end{remark}

\begin{remark}\label{pf}
Many existing works on MTSED use the DC power flow model as a constraint to describe active power balance~at each bus (e.g., \cite{dis4,gm,dc}), but it cannot sufficiently handle voltage limits that are critical to satisfactory performance of power networks.  Instead, problem \eqref{op} adopts the DLPF model developed in \cite{lpf}.  Compared with DC flow, the DLPF model is a good approximation of AC power flow \cite{lpf},~and allows MTSED to take both the constraints on reactive power balance as well as bus voltages into account.  Moreover, DLPF adds bus reactive power injections into the optimization problem as decision variables, and the generation cost function is irrespective of reactive flows, which makes the resulting problem only generic convex on reactive power rather than strictly/strongly convex. Based on [14],  designing a distributed algorithm for solving problems with both generic convex objective functions and hard feasibility constraints is still a challenging problem in both the power and optimization communities.
\end{remark}

\section{Main Results}

In this section, we propose a projection-based algorithm to solve  \eqref{op} in a distributed manner and analyze the optimality as well as asymptotic convergence of the designed method.

\subsection{Preliminaries}

Firstly, we provide some necessary preliminaries that will be used throughout this section.  Let $\xi$ be a vector in $\mathbb{R}^n$, and $S\subset\mathbb{R}^n$ be a closed convex set. Then, we denote $\text{P}_{S}(\xi)=\text{arg\hspace{1pt}min}_{\eta\in S}\Vert \xi-\eta\Vert$ as the projection of $\xi$ onto $S$ and  $\text{C}_{S}(\xi)=\{\rho\in$ $\mathbb{R}^n|\rho^T(\eta-\xi)\leq 0,\forall\eta\in S\}$ as the normal cone of $S$ at $\xi$. The following four lemmas give some basic properties of the projection operator and normal cone.

\begin{lemma}[\cite{sf}]\label{lemma1}
For any two vectors $\xi,\eta\in\mathbb{R}^n$, $\text{P}_{\mathbb{R}_+^n}(\xi+\eta)=\xi$ if and only if $\xi\geq 0_{n\times1}$, $\eta\leq 0_{n\times1}$, $\xi^T\eta=0$.
\end{lemma}
\begin{lemma}[\cite{no}]\label{lemma2}
Let $S\subset\mathbb{R}^n$ be a closed convex set and $\xi\in S$, then $\rho\in \text{C}_S(\xi)$ if and only if $\text{P}_{S}(\xi+\rho)=\xi$.
\end{lemma}

\begin{lemma}[\cite{sf}]\label{lemma3}
Let $S\subset$ $\mathbb{R}^n$ be a closed convex set. Define $\psi(\xi,\eta)=\frac{1}{2}(\Vert\xi-\eta\Vert^2-\Vert\xi-\text{P}_S(\xi)\Vert^2)$ with $\xi\in\mathbb{R}^n$, $\eta\in S$. Then, $\psi(\xi,\eta)$ is continuously differentiable on $\xi$ with $\partial\psi/\partial\xi=\text{P}_{S}(\xi)-\eta$ and satisfies the following inequality 
\begin{equation}
\psi(\xi,\eta)\geq\frac{1}{2}\Vert \text{P}_{S}(\xi)-\eta\Vert^2.
\end{equation}
 \end{lemma}
 
 \begin{lemma}[\cite{the}]\label{lemma4}
Let $S\subset\mathbb{R}^n$ be a closed convex set, then
\begin{equation}\label{pr1}
(\text{P}_S(\xi)-\eta)^T(\xi-\text{P}_{S}(\xi))\geq 0,~\forall \xi\in\mathbb{R}^n,\eta\in S.
\end{equation}
 \end{lemma}

Let $x(t)$ be a solution of an autonomous system 
\begin{equation}\label{nlsys}
\dot{x}=f(x)
\end{equation} 
with $f(\cdot):D\to\mathbb{R}^n$ being a locally Lipschitz map from a domain $D\subset\mathbb{R}^n$ into $\mathbb{R}^n$. 
\begin{definition} [\cite{ns}]
A point $x^*$ is said to be a positive limit point of $x(t)$ if there is a sequence $\{t_n\}$ with $t_n\to+\infty$ as $n\to+\infty$ such that $x(t_n)\to x^*$ as $n\to+\infty$. 
\end{definition}
\begin{definition} [\cite{ns}]\label{lemma5}
 The set of all positive limit points of $x(t)$ is called the positive limit set of $x(t)$. 
 \end{definition}

The following lemma presents a fundamental property of positive limit sets

\begin{lemma}[\cite{ns}]\label{lem5}
If a solution $x(t)$ of system \eqref{nlsys} is bounded and belongs to $D$ for $t\geq 0$, then its positive limit set $\mathcal{L}^+$ is nonempty, compact, and invariant. Moreover, $x(t)$ approaches $\mathcal{L}^+$ as $t\to+\infty$.
\end{lemma}


\subsection{Distributed Projection-Based Algorithm}

To solve problem \eqref{op} in a distributed way, we assume there exists a communication network that has the same topology as that of the physical grid and suppose that any two directly interconnected buses in the communication network can share information with each other. These two assumptions are widely used in the literature of distributed ED (e.g., \cite{15,dis4,gm}), since communication between neighboring buses in power networks can be easily achieved by technologies such as  802.2.15.4/ZigBee in practice\cite{huawei}.

At each time $t=hT_o$, $h\in\mathbb{Z}_+$, we run the following dynamic system at each bus $i$, $i\in\mathcal{N}$, to solve the optimal solution of \eqref{op} 
\begin{subequations}\label{al}
\begin{align}
\dot{p}_{gi}=&\tilde p_{gi}-p_{gi}-a_{gi}\tilde p_{gi}-b_{gi}1_{\tau}-\lambda_{pi}-\rho_{pi}\nonumber\\
&+(H_g-I_{\tau})(\tilde{\mu}_{Mi}-\tilde{\mu}_{mi})\label{al1}\\
\dot{q}_{gi}=&\tilde q_{gi}-q_{gi}-\lambda_{qi}-\rho_{qi}\label{al2}\\
\dot{p}_{ci}=&\tilde p_{ci}-p_{ci}-a_{si}1_{\tau}+\lambda_{pi}+\rho_{pi}-\eta_{ci}T_oH_s\nonumber\\
&\times(\tilde{\gamma}_{Mi}-\tilde{\gamma}_{mi})\label{al3}\\
\dot{p}_{di}=&\tilde p_{di}-p_{di}-a_{si}1_\tau-\lambda_{pi}-\rho_{pi}+\eta_{di}^{-1}T_oH_s\nonumber\\
&\times(\tilde{\gamma}_{Mi}-\tilde{\gamma}_{mi})\label{al4}\\
\dot{v}_{i}=&\tilde v_{i}-v_{i}+\sum_{j=1}^n(g_{ij}(\lambda_{pj}+\rho_{pj})-b_{ij}(\lambda_{qj}+\rho_{qj}))\label{al5}\\
\dot{\theta}_{i}=&-\sum_{j=1}^n(b_{ij}'(\lambda_{pj}+\rho_{pj})+g_{ij}(\lambda_{qj}+\rho_{qj}))\label{al6}\\
%
\dot{\lambda}_{pi}&=\tilde{\lambda}_{pi}, ~~~~~~~~~~~~~~\dot{\lambda}_{qi}=\tilde{\lambda}_{qi}\label{al7}\\
\dot{\rho}_{pi}&=-\rho_{pi}+\tilde{\lambda}_{pi},~~~\dot{\rho}_{qi}=-\rho_{qi}+\tilde{\lambda}_{qi}\label{alrho}\\
\dot{\mu}_{Mi}&=\tilde{\mu}_{Mi}-\mu_{Mi},~~\dot{\mu}_{mi}=\tilde{\mu}_{mi}-{\mu_{mi}}\label{al9}\\
\dot{\gamma}_{Mi}&=\tilde{\gamma}_{Mi}-\gamma_{Mi},~~
\dot{\gamma}_{mi}=\tilde{\gamma}_{mi}-{\gamma_{mi}}\label{al11}
%
\end{align}
\end{subequations}
where $p_{gi}=\text{col}({ p}_{gi1},\dots,{p}_{gi\tau})\in\mathbb{R}^{\tau}$ and vectors ${q}_{gi}$, $p_{ci}$, ${p}_{di}$, ${v}_i$, $\theta_i$, $\lambda_{pi}$, $\lambda_{qi}$, ${\rho}_{pi}$,  ${\rho}_{qi}$,  ${\mu}_{Mi}$, ${\mu}_{mi}$, ${\gamma}_{Mi}$, ${\gamma}_{mi}$, $\tilde p_{gi}$, $\tilde q_{gi}$, $\tilde p_{ci}$, $\tilde p_{di}$, $\tilde v_{i}\in\mathbb{R}^{\tau}$ are defined similarly as $p_{gi}$ with $ p_{gik}$, $ q_{gik}$, $ p_{cik}$, $ p_{dik}$, $ v_{ik}$, $\theta_{ik}$, $\lambda_{pik}$, $\lambda_{qik}$, $\rho_{pik}$, $\rho_{qik}$, $\mu_{Mik}$, ${\mu}_{mik}$, ${\gamma}_{Mik}$, ${\gamma}_{mik}$, $\forall k\in\mathcal{T}$ being the state variables of system \eqref{al} at bus $i$. Further, $\tilde p_{gi}$, $\tilde q_{gi}$, $\tilde p_{ci}$, $\tilde p_{di}$, $\tilde v_{i}$ are projections of $ p_{gi}$, $ q_{gi}$, $ p_{ci}$, $ p_{di}$, $ v_{i}$ defined as
\begin{subequations}
\begin{align}
\tilde p_{gi}=&\text{P}_{\Omega_{pi}}(p_{gi}),~\tilde q_{gi}=\text{P}_{\Omega_{qi}}(q_{gi}),~\tilde p_{ci}=\text{P}_{\Omega_{ci}}(p_{ci})\label{al12}\\
\tilde p_{di}=&\text{P}_{\Omega_{di}}(p_{di}),~\tilde v_{i}=\text{P}_{\Omega_{vi}}(v_{i}).\label{al13}
\end{align}
\end{subequations}
where sets $\Omega_{pi},\Omega_{qi},\Omega_{ci},\Omega_{di},\Omega_{vi}\subset\mathbb{R}^{\tau}$ are defined by $\Omega_{pi}=[\underline{p}_{gi},\overline{p}_{gi}]\times\dots\times[\underline{p}_{gi},\overline{p}_{gi}]$, $\Omega_{qi}=[\underline{q}_{gi},\overline{q}_{gi}]\times\dots\times[\underline{q}_{gi},\overline{q}_{gi}]$, $\Omega_{ci}=[\underline{q}_{ci},\overline{q}_{ci}]\times\dots\times[\underline{q}_{ci},\overline{q}_{ci}]$, $\Omega_{di}=[\underline{q}_{di},\overline{q}_{di}]\times\dots\times[\underline{q}_{di},\overline{q}_{di}]$ and $\Omega_{vi}=[\underline{v}_{i},\overline{v}_{i}]\times\dots\times[\underline{v}_{i},\overline{v}_{i}]$. Variables $\tilde{\lambda}_{pi}$, $\tilde{\lambda}_{qi}$, $\tilde{\mu}_{Mi}$, $\tilde{\mu}_{mi}$, $\tilde{\gamma}_{Mi}$, $\tilde{\gamma}_{mi}$ are introduced to simplify expressions of  equations in \eqref{al} and are defined as
\begin{subequations}\label{con1}
\begin{align}
\tilde{\lambda}_{pi}=&\tilde p_{gi}-d_{pi}-\tilde p_{ci}+\tilde p_{di}-\sum_{j=1}^n(g_{ij}\tilde v_{j}-b_{ij}'\theta_{j})\label{con1a}\\
\tilde{\lambda}_{qi}=&\tilde q_{gi}-d_{qi}+\sum_{j=1}^n(b_{ij}\tilde v_{j}+g_{ij}\theta_{j})\label{con1b}\\
\tilde{\mu}_{Mi}=&\text{P}_{\mathbb{R}_{+}^{\tau}}({\mu}_{Mi}+(I_\tau-H_g^T)\tilde p_{gi}-p_{gi}[0]h_{g0}-\overline r_{i}T_o1_\tau)\label{con1c}\\
\tilde{\mu}_{mi}=&\text{P}_{\mathbb{R}_{+}^{\tau}}({\mu}_{mi}+\underline r_{i}T_o1_\tau- (I_\tau-H_g^T)\tilde p_{gi}+p_{gi}[0]h_{g0})\label{con1d}\\
\tilde{\gamma}_{Mi}=&\text{P}_{\mathbb{R}_{+}^{\tau}}({\gamma}_{Mi}+(c_{i}[0]-\overline c_{i})1_{\tau}+\eta_{ci}T_oH_s^T\tilde p_{ci}\notag\\
&-\eta^{-1}_{di}T_oH_s^T\tilde p_{di})\label{con1e}\\
\tilde{\gamma}_{mi}=&\text{P}_{\mathbb{R}_{+}^{\tau}}({\gamma}_{mi}+(\underline c_{i}-c_{i}[0])1_\tau-\eta_{ci}T_oH_s^T\tilde p_{ci}\notag\\
&+\eta^{-1}_{di}T_oH_s^T\tilde p_{di})\label{con1f}
\end{align}
\end{subequations}
where constant vectors $d_{pi},d_{qi},h_{g0}\in\mathbb{R}^{\tau}$ are defined by~$d_{pi}=\text{col}(d_{pi}[1],\dots,d_{pi}[\tau])$, $d_{qi}=\text{col}(d_{qi}[1],\dots,d_{qi}[\tau])$, $h_{g0}=$ $\text{col}(1,0_{(\tau-1)\times1})$, and matrices $H_{g},H_s\in\mathbb{R}^{\tau\times\tau}$ are defined~by
\begin{align}
H_g&=
\left(
\begin{array}{cccc}
0_{(\tau-1)\times 1}&I_{\tau-1} \\  
0&0_{1\times(\tau-1)}
\end{array}
\right),H_s=
\left(
\begin{array}{cccc}
1&1&\cdots&1 \\  
0&1&\cdots&1\\ 
\vdots &\vdots&\ddots&\vdots\\ 
0&0&\cdots&1
\end{array}
\right).\nonumber
\end{align}
In what follows, we will use the term algorithm  \eqref{al} or system \eqref{al} interchangeably.

Variables $\tilde p_{gik}$, $\tilde q_{gik}$, $\tilde p_{cik}$, $\tilde p_{dik}$, $\tilde v_{ik}$, $ \theta_{ik}$ are used to solve the optimal solution $p^*_{gi}[k]$, $q^*_{gi}[k]$, $p^*_{ci}[k]$, $p^*_{di}[k]$, $v^*_{i}[k]$, $\theta^*_{i}[k]$, $\forall i\in\mathcal{N},k\in\mathcal{T}$ of the MTSED problem. We will show in the next subsection that when system \eqref{al} is at its steady-state condition, the steady-state values of $\tilde p_{gik}$, $\tilde q_{gik}$, $\tilde p_{cik}$, $\tilde p_{dik}$, $\tilde v_{ik}$, $ \theta_{ik}$, $\forall i\in\mathcal{N},k\in\mathcal{T}$ fulfil the feasibility and optimality conditions of~\eqref{op}. So, if the equilibrium point of \eqref{al} is asymptotically convergent, we can run system  \eqref{al}  at each bus $i$ to find out the optimal solution of the MTSED problem. Furthermore, if there is no direct connection between two buses $i,j\in\mathcal{N}$, we have $g_{ij}=b_{ij}=b_{ij}'=0$ \cite{lpf}, which means that system \eqref{al} only depends on local information at each bus and information $v_j$, $\theta_j$, $\lambda_{pj}$, $\lambda_{qj}$, from its neighbors.  Therefore, problem \eqref{op} can be solved in a distributed way,  and the bus privacy is also preserved as each bus has the autonomy and authority to formulate its own objective function as well as feasibility set while implementing the designed method.

In particular, subsystem \eqref{al1} is used to compute $p_{gik}$, the unconstrained version of $\tilde p_{gik}$,  where $\tilde p_{gik}-p_{gik}$ is used to measure the difference between the constrained and unconstrained active power dispatches of SGs;  $\lambda_{pik}$ is used to estimate the SG's incremental cost $a_{gi}\tilde p_{gik}+b_{gi}$, i.e., the first derivative of the generation cost function; $\rho_{pik}$ is introduced~to ensure the algorithm convergence (see Theorem 3 for details) and does not affect the equilibrium point of \eqref{al} as it is equal~to 0 at the steady state; and the remaining term consists of information of $\mu_{Mik}$, $\mu_{mik}$ from \eqref{al9}  that are used to make the active power outputs of SGs satisfy the ramping limits.

Similar to \eqref{al1}, we use subsystems \eqref{al2}-\eqref{al4} to compute the unconstrained versions of $\tilde{q}_{gik}$, $\tilde p_{cik}$ and $\tilde p_{dik}$, respectively. Different from $\lambda_{pik}$ in \eqref{al1}, the estimated incremental cost ${\lambda}_{qik}$ related to reactive power generation in \eqref{al2} needs to approach 0 as the generation cost functions are independent of reactive power outputs. Subsystems \eqref{al3}, \eqref{al4} employ~the feedback information of $\gamma_{Mik}$, $\gamma_{mik}$ from  \eqref{al11} that aim to make sure that the charging/discharging powers of ESDs fulfil the energy capacity constraints. 

Subsystems \eqref{al5}, \eqref{al6} are adopted to make all estimated incremental costs ${\lambda}_{pik}$, ${\lambda}_{qik}$ satisfy their corresponding optimality conditions at the equilibrium point. Further, subsystems \eqref{al7}, \eqref{alrho} are used to enforce the active and reactive power balance constraints. 

Finally, equations \eqref{al12}, \eqref{al13} are used to make $\tilde p_{gik}$, $\tilde q_{gik}$, $\tilde p_{cik}$, $\tilde p_{dik}$, $\tilde v_{ik}$ fulfil the feasibility constraints \eqref{c3}, \eqref{c4}, \eqref{c6}, \eqref{c7} and \eqref{c10} by projecting  $p_{gik}$, $q_{gik}$, $p_{cik}$, $p_{dik}$, $v_{ik}$ into the corresponding feasible sets.



\begin{remark}
It should be noted that the optimization algorithms in some existing works (e.g., \cite{ca2,dis1,dis2,dis3,dis4,yiguang}) depend on the assumptions of the strict/strong convexity of objective functions and strong Slater condition. As pointed out in \cite{gclc}, the design of distributed algorithms for an optimization problem with generic convex objective functions is challenging. In this paper, a linear cost function is used~to~quantify~the~operation costs of ESDs, which makes the objective function of \eqref{op} is only convex~rather~than~strictly/strongly~convex. From this point of view,  the obtained results complements those early works (e.g. \cite{ca2,dis1,dis2,dis3,dis4,yiguang}) by removing the restriction of the strict/strong convexity on objective functions. 
\end{remark}
 
\begin{remark}
Like \cite{dis4,gm,dc,yiguang}, the physical constraints on transmission line flows that are generally important to~be~included in the formulation of ED problems are not considered in the present paper for simplicity. It is worth pointing out~that, similar to generator ramping limits and storage energy limits, line flow constraints are also convex constraints on the decision variables of the optimization problem \eqref{op} particularly pertaining to bus phase angles and voltage magnitudes.  Hence, in principle, the proposed algorithm \eqref{al} for  \eqref{op} can be easily extended to solve problems subject to line power constraints. We leave this topic for future work.
 \end{remark}

\begin{remark}
Inspired by \cite{gclc}, we introduce variables $\rho_{pik}$, $\rho_{qik}$, $i\in\mathcal{N},k\in\mathcal{T}$ into the designed algorithm \eqref{al}. They serve as the phase lead compensator and are used to guarantee that the distributed algorithm \eqref{al} converges to the optimal solution of \eqref{op} with  generic convexity. In \cite{gm}, we consider a similar problem as \eqref{op}. The main differences  of the current paper from \cite{gm} are threefold. Firstly, we consider both active power balance and bus voltage constraints in \eqref{op}, whereas \cite{gm} only considers active power~balance~constraint. Secondly, we use different distributed algorithms  where in the current paper we introduce variables $\rho_{pik}$, $\rho_{qik}$ in the designed algorithm but does not in that of \cite{gm}. Finally, we provide the  theoretical proofs of the asymptotic convergence of the designed distributed  algorithm, whereas only the corresponding simulation study is given in \cite{gm}. 
\end{remark}

%

\subsection{Optimality of the Distributed Algorithm}

In this subsection, we  establish the relationship between the optimal solution of the MTSED problem \eqref{op} and equilibrium point of the dynamical system \eqref{al}, which is stated as follows. 

\begin{theorem}\label{theorem2}
Let $p_{gik}^*$, $q_{gik}^*$, $p^*_{cik}$, $p^*_{dik}$, $v^*_{ik}$, $\theta^*_{ik}$, $\lambda^*_{pik}$, $\lambda^*_{qik}$, $\rho^*_{pik}$, $\rho^*_{qik}$, $\mu^*_{Mik}$, ${\mu}^*_{mik}$, ${\gamma}^*_{Mik}$, ${\gamma}^*_{mik}$, $\forall i\in\mathcal{N},k\in\mathcal{T}$ be an equilibrium point of system \eqref{al}, then the corresponding $\tilde{p}^*_{gik}$, $\tilde{q}^*_{gik}$, $\tilde{p}^*_{cik}$, $\tilde{p}^*_{dik}$, $\tilde{v}^*_{ik}$, $\theta^*_{ik}$, $\forall i\in\mathcal{N},k\in\mathcal{T}$ are an optimal solution of the MTSED problem \eqref{op}.
\end{theorem}

\textit{Proof: i) feasibility.} Clearly, $\tilde p_{gik}^*$, $\tilde q_{gik}^*$, $\tilde p_{cik}^*$, $\tilde p^*_{dik}$, $\tilde v^*_{ik}$, $\theta_{ik}^*$ satisfy \eqref{c1}, \eqref{c2} by setting $\dot{{\lambda}}_{pik}=\dot{{\lambda}}_{qik}=0$ in \eqref{al7}  at the equilibrium point. Moreover, they also satisfy \eqref{c3}, \eqref{c4}, \eqref{c6}, \eqref{c7}, \eqref{c10} because projections in \eqref{al12} and \eqref{al13}.

Setting $\dot{\mu}_{Mik}=\dot{\mu}_{mik}=0$ in \eqref{al9} and \eqref{al9} gives
\begin{equation}
\label{mu0}
\begin{split}
\mu_{Mik}^*=&\text{P}_{\mathbb{R}_+}(\mu_{Mik}^*+(\tilde p_{gik}^*- \tilde p^*_{gi(k-1)})-\overline r_{i}T_o)\\
\mu^*_{mik}=&\text{P}_{\mathbb{R}_+}(\mu^*_{mik}+\underline r_{i}T_o-(\tilde p_{gik}^*-\tilde p_{gi(k-1)}^*)).
\end{split}
\end{equation}
According to Lemma \ref{lemma1}, equations in \eqref{mu0} are equivalent to
\begin{subequations}\label{mumu}
\begin{align}
&{\mu}_{Mik}^*\geq 0,~{\mu}_{mik}^*\geq 0\label{mumu1}\\
&\overline r_{i}T_o\geq\tilde p_{gik}^*- \tilde p_{gi(k-1)}^*\geq \underline r_{i}T_o\label{mumu2}\\
&\mu^*_{Mik}((\tilde p^*_{gik}-\tilde p^*_{gi(k-1)})-\overline r_{i}T_o)=0\label{mumu3}\\
&\mu^*_{mik}(\underline r_{i}T_o-(\tilde p^*_{gik}-\tilde p^*_{gi(k-1)}))=0\label{mumu4}
\end{align}
\end{subequations}
where \eqref{mumu2} ensures \eqref{c5} is satisfied.  Similarly, setting $\dot{{\gamma}}_{Mik}=\dot{{\gamma}}_{mik}=0$ in  \eqref{al11} and applying Lemma \ref{lemma1} give
\begin{subequations}\label{gamma0}
\begin{align}
&{\gamma}_{Mik}^*\geq 0,~{\gamma}_{mik}^*\geq 0\label{gamma1}\\
&\underline c_{i}\leq c_{i}[0]+T_o\sum_{l=1}^{k}(\eta_{ci}\tilde p_{cil}-\eta_{di}^{-1}\tilde p_{dil})\leq\overline c_{i}\label{gamma2}\\
&\gamma^*_{Mik}(c_{i}[0]+T_o\sum_{l=1}^{k}(\eta_{ci}\tilde p_{cil}-\eta_{di}^{-1}\tilde p_{dil})-\overline c_{i})=0\label{gamma3}\\
&\gamma^*_{mik}(\underline c_{i}-c_{i}[0]-T_o\sum_{l=1}^k(\eta_{ci}\tilde p_{cil}-\eta^{-1}_{di}\tilde p_{dil}))=0\label{gamma4}
\end{align}
\end{subequations}
where \eqref{gamma2} guarantees that \eqref{c9} is satisfied.  Therefore, $\tilde p_{gik}^*$, $\tilde q_{gik}^*$, $\tilde p_{cik}^*$, $\tilde p^*_{dik}$, $\tilde v^*_{ik}$, $\theta_{ik}^*$, $\forall i\in\mathcal{N},k\in\mathcal{T}$ satisfy all constraints in \eqref{op}, and thus are a feasible solution of \eqref{op}.

{\textit{ii) optimality.} \eqref{mumu1}, \eqref{mumu3}, \eqref{mumu4}, \eqref{gamma1}, \eqref{gamma3}, \eqref{gamma4} enforce the dual feasibility and complementary slackness of the KKT conditions of the MTSED problem \eqref{op} \cite{no}.

From $\dot{{\lambda}}_{pik}=\dot{{\rho}}_{pik}=0$ in \eqref{al7}-\eqref{alrho} at the equilibrium point, we have $\rho_{pik}^*=0$. Setting $ \dot{p}_{gik}=0$ in \eqref{al1} gives
\begin{equation}
\begin{split}
p_{gik}^*=&\tilde p_{gik}^*-a_{gi}\tilde p^*_{gik}-b_{gi}-\lambda_{pik}^*+(\mu^*_{Mi(k+1)}\\
&-\mu^*_{mi(k+1)})-(\mu^*_{Mik}-\mu^*_{mik})\label{p}
\end{split}
\end{equation}
where we use \eqref{mu0} to get \eqref{p}. Then, applying the projection operator $\text{P}_{[\underline{p}_{gi},\overline{p}_{gi}]}(\cdot)$ on both sides of \eqref{p} gives
\begin{align}\label{14}
\tilde{p}_{gik}^*=&\text{P}_{[\underline{p}_{gi},\overline{p}_{gi}]}(p_{gik}^*)\notag\\
=&\text{P}_{[\underline{p}_{gi},\overline{p}_{gi}]}(\tilde p_{gik}^*-a_{gi}\tilde p^*_{gik}-b_{gi}-\lambda_{pik}^*\\
&+(\mu^*_{Mi(k+1)}-\mu^*_{mi(k+1)})-(\mu^*_{Mik}-\mu^*_{mik}))\notag
\end{align}
According to Lemma \ref{lemma2}, equation \eqref{14} is equivalent to
\begin{equation}
\begin{split}
0\in&a_{gi}\tilde p^*_{gik}+b_{gi}+\lambda_{pik}^*-(\mu^*_{Mi(k+1)}-\mu^*_{mi(k+1)})\\
&+(\mu^*_{Mik}-\mu^*_{mik})+\text{C}_{[\underline{p}_{gi},\overline{p}_{gi}]}(\tilde p_{gik}^*)
\end{split}
\end{equation}
for all $i\in\mathcal{N}$ and $k\in\mathcal{T}$. Therefore,  the KKT stationarity conditions of \eqref{op} with respect to $\tilde p^*_{gik}$ is satisfied.  Similarly, $\dot{q}_{gik}=0$, $\dot{p}_{cik}=0$, $\dot{p}_{dik}=0$,  $\dot{v}_{ik}=0$, $\dot{\theta}_{ik}=0$ at the equilibrium point give that the stationarity of KKT conditions with respect to other decision variables of \eqref{op} are satisfied. 

As a result, the feasible solutions $\tilde{p}^*_{gik}$, $\tilde{q}^*_{gik}$, $\tilde{p}^*_{cik}$, $\tilde{p}^*_{dik}$, $\tilde{v}^*_{ik}$, $\theta^*_{ik}$, $\forall i\in\mathcal{N},k\in\mathcal{T}$ satisfy all optimality conditions, and are certainly an optimal solution of problem \eqref{op}  \cite{no}. $\hfill\blacksquare$ 

\begin{figure*}[b]
\normalsize
\hrulefill
\setcounter{equation}{19}
\begin{equation}
\begin{split}
%
%
%
\mathcal{C}&=
\left(
\begin{array}{cccccccccc}
I_{n\tau}&0_{n\tau\times n\tau}&-I_{n\tau}&I_{n\tau}&-G\otimes I_\tau&B'\otimes I_\tau \\
0_{n\tau\times n\tau}&I_{n\tau}&0_{n\tau\times n\tau}&0_{n\tau\times n\tau}&B\otimes I_{\tau}&G\otimes I_\tau
\end{array}
\right)\\
%
%
\mathcal{E}&=
\left(
\begin{array}{cccccccc}
I_{n\tau}-I_n\otimes H_g^T&0_{n\tau\times n\tau}&0_{n\tau\times n\tau}&0_{n\tau\times n\tau}&0_{n\tau\times n\tau}&0_{n\tau\times n\tau} \\
I_n\otimes H_g^T-I_{n\tau}&0_{n\tau\times n\tau}&0_{n\tau\times n\tau}&0_{n\tau\times n\tau}&0_{n\tau\times n\tau}&0_{n\tau\times n\tau}\\
0_{n\tau\times n\tau}&0_{n\tau\times n\tau}&T_o(\Gamma_c\otimes H_s^T)&-T_o(\Gamma^{-1}_d\otimes H_s^T)&0_{n\tau\times n\tau}&0_{n\tau\times n\tau}\\
0_{n\tau\times n\tau}&0_{n\tau\times n\tau}&-T_o(\Gamma_c\otimes H_s^T)&T_o(\Gamma^{-1}_d\otimes H_s^T)&0_{n\tau\times n\tau}&0_{n\tau\times n\tau}
\end{array}
\right)\\
%
%
\label{matrix}
\end{split}
\end{equation}
\setcounter{equation}{17}
\end{figure*}

\subsection{Convergence Analysis}
In this subsection, we analyse asymptotic convergence of system \eqref{al}.  Let $p_{g}=\text{col}(p_{g1},\dots,p_{gn})\in\mathbb{R}^{n\tau}$, and define vectors $q_g$, ${p}_c$, ${p}_d$, ${v}$, $\theta$, $\lambda_p$, $\lambda_q$, ${\rho}_p$, ${\rho}_q$, ${\mu}_M$, ${\mu}_m$, ${\gamma}_M$, ${\gamma}_m$ $\tilde{p}_g$, $\tilde{q}_g$, $\tilde{p}_c$, $\tilde{p}_d$, $\tilde{v}\in\mathbb{R}^{n\tau}$ similarly as $p_{g}$. Then, system \eqref{al} can be rewritten as follows
\begin{equation}\label{vector}
\begin{split}
\dot{p}_{g}=&-p_{g}+(I_{n\tau}-A_{g}\otimes I_\tau)\tilde p_{g}-b_{g}\otimes 1_\tau-\lambda_{p}\\
&-\rho_{p}+(I_n\otimes H_g-I_{n\tau})(\tilde{\mu}_M-\tilde{\mu}_m)\\
%
%
\dot{q}_{g}=&-q_{g}+\tilde q_{g}-\lambda_{q}-\rho_q\\
\dot{p}_{c}=&-p_{c}+\tilde p_{c}-a_{s}\otimes 1_\tau+\lambda_{p}+\rho_p\\
&-T_o(\Gamma_c\otimes H_s)(\tilde{\gamma}_{M}-\tilde{\gamma}_{m})\\
%
%
%
\dot{p}_{d}=&-p_{d}+\tilde p_{d}-a_{s}\otimes 1_\tau-\lambda_{p}-\rho_p\\
&+T_o(\Gamma^{-1}_d\otimes H_s)(\tilde{\gamma}_{M}-\tilde{\gamma}_{m})\\
%
%
\dot{v}=&-v+\tilde{v}+(G\otimes I_\tau)(\lambda_{p}+\rho_{p})\\
&-(B\otimes I_\tau)(\lambda_{q}+\rho_{q})\\
\dot{\theta}=&-(B'\otimes I_\tau)(\lambda_{p}+\rho_{p})-(G\otimes I_\tau)(\lambda_{q}+\rho_{q})\\
\dot{ \lambda}_{p}=&\tilde{\lambda}_{p},~\dot{\lambda}_{q}=\tilde{\lambda}_{q}\\
\dot{\rho}_{p}=&-\rho_p+\tilde{\lambda}_{p},~\dot{\rho}_{q}=-\rho_q+\tilde{\lambda}_{q}\\
\dot{\mu}_{M}=&\tilde{\mu}_{M}-\mu_{M},~~\dot{\mu}_{m}=\tilde{\mu}_{m}-{\mu_{m}}\\
\dot{\gamma}_{M}=&\tilde{\gamma}_{M}-\gamma_{M},~~\dot{\gamma}_{m}=\tilde{\gamma}_{m}-{\gamma_{m}}
%
\end{split}
\end{equation}
where $\tilde p_{g}=\text{P}_{\Omega_p}(p_{g})$, $\tilde q_{g}=\text{P}_{\Omega_q}(q_{g})$, $\tilde p_{c}=\text{P}_{\Omega_c}(p_{c})$, $\tilde p_{d}=\text{P}_{\Omega_d}(p_{d})$, $\tilde v=\text{P}_{\Omega_v}(v)$ with $\Omega_{p}=\Omega_{p1}\times\dots\times$ $\Omega_{pn}$, $\Omega_{q}=\Omega_{q1}\times\dots\times\Omega_{qn}$, $\Omega_{c}=\Omega_{c1}\times\dots\times\Omega_{cn}$, $\Omega_{d}=\Omega_{d1}\times\dots\times\Omega_{dn}$, $\Omega_{v}=\Omega_{v1}\times\dots\times\Omega_{vn}\subset\mathbb{R}^{n\tau}$; and $\tilde{\lambda}_{p}=\tilde p_{g}-d_{p}-\tilde p_{c}+\tilde p_{d}-(G\otimes I_\tau)\tilde v+(B'\otimes I_\tau)\theta$, $\tilde{\lambda}_{q}=\tilde q_{g}-d_{q}+(B\otimes I_\tau)\tilde v+(G\otimes I_\tau)\theta$, 
$\tilde{\mu}_{M}=\text{P}_{\mathbb{R}_+^{n\tau}}\left((I_{n\tau}-I_n\otimes H_g^T)\tilde p_{g}-p_{g0}\otimes h_{g0}-T_o(\overline r\otimes 1_\tau)\right)$, 
$\tilde{\mu}_{m}=\text{P}_{\mathbb{R}_+^{n\tau}}\left(T_o(\underline r\otimes 1_\tau)-(I_{n\tau}-I_n\otimes H_g^T)\tilde p_{g}+p_{g0}\otimes h_{g0}\right)$, 
$\tilde{\gamma}_{M}=\text{P}_{\mathbb{R}_+^{n\tau}}\left((c_{0}-\overline c)\otimes 1_\tau+T_o(\Gamma_{c}\otimes H_s^T)\tilde p_{c}-T_o(\Gamma^{-1}_{d}\right.$ $\left. \otimes H_s^T)\tilde p_{d}\right)$ 
and $\tilde{\gamma}_{m}=\text{P}_{\mathbb{R}_+^{n\tau}}\left((\underline c-c_0)\otimes 1_\tau-T_o(\Gamma_{c}\otimes H_s^T)\tilde p_{c}\right.$ $\left.+T_o(\Gamma^{-1}_{d}\otimes H_s^T)\tilde p_{d}\right)$.



Constant vectors $d_p,d_q\in\mathbb{R}^{n\tau}$, $a_s,b_g,p_{g0},\underline{r},\overline{r},c_{0},\underline{c},\overline{c}\in\mathbb{R}^{n}$ are defined by $d_p=\text{col}(d_{p1},\dots,d_{pn})$, $d_q=\text{col}(d_{q1},\dots,$ $d_{qn})$ with $d_{pi},d_{qi}\in\mathbb{R}^{\tau}$, $\forall i\in\mathcal{N}$ defined in \eqref{con1}; $a_{s}=\text{col}(a_{s1},$ $\dots,a_{sn})$, $b_g=\text{col}(b_{g1},\dots,b_{gn})$, $p_{g0}=\text{col}(p_{g1}[0],\dots,$ $p_{gn}[0])$, $\underline{r}=\text{col}(\underline{r}_1,\dots,\underline{r}_n)$, $\overline{r}=\text{col}(\overline{r}_1,\dots,\overline{r}_n)$, $c_0=\text{col}(c_{1}[0],\dots,c_{n}[0])$, $\underline{c}=\text{col}(\underline{c}_1,\dots,\underline{c}_n)$, and $\overline{c}=\text{col}(\overline{c}_1,\dots,$ $\overline{c}_n)$. Furthermore, matrices $A_g,\Gamma_c,\Gamma_d\in\mathbb{R}^{n\times n}$ are defined by $A_g=\text{diag}(a_{g1},\dots,a_{gn})$, $\Gamma_{c}=\text{diag}(\eta_{c1},\dots,\eta_{cn})$, and $\Gamma_{d}=$ $\text{diag}(\eta_{d1},\dots,\eta_{dn})$.

Let $x=\text{col}(p_g,q_g,p_c,p_d,v,\theta)$, $\tilde{x}=\text{col}(\tilde {{p}}_g,$ $\tilde{ q}_g,\tilde{{p}}_c,\tilde{p}_d,\tilde{v},\theta)$, $y=\text{col}(\lambda_p,\lambda_q)$, $z=\text{col}(\mu_M,\mu_m,\gamma_M,\gamma_m)$, $\rho=\text{col}(\rho_p,\rho_q)$, and  $\Omega=\Omega_p\times$ $\Omega_q\times\Omega_c\times\Omega_d\times\Omega_v\times\mathbb{R}_{n\tau}$. Then,  system \eqref{vector} can be re-expressed in a compact form as 
\begin{equation}\label{eal}
\begin{split}
\dot{ x}&=- x+\tilde x-\mathcal{A}\tilde x-\mathcal{B}-\mathcal{C}^T(y+\rho)-\mathcal{E}^Tz^+\\
\dot{y}&=\mathcal{C}\tilde x-\mathcal{D}\\
\dot{z}&=z^+-z\\
\dot{\rho}&=-\rho+\mathcal{C}\tilde x-\mathcal{D}
\end{split}
\end{equation}
where $\tilde x=\text{P}_\Omega(x)$, $z^+=\text{P}_{\mathbb{R}_{+}^{4n\tau}}(\mathcal{E}\tilde x-\mathcal{F}+z)$, $\mathcal{A}=\text{diag}(A_g\otimes I_\tau,0_{5n\tau\times 5n\tau})$, $\mathcal{B}=\text{col}(b_g\otimes 1_\tau,0_{n\tau\times 1},a_s\otimes 1_\tau,a_s\otimes 1_\tau,0_{2n\tau\times 1})$, $\mathcal{D}=\text{col}(d_p,d_q)$, $\mathcal{F}=\text{col}(p_{g0}\otimes h_{g0}+T_o(\overline r\otimes 1_\tau),- p_{g0}\otimes h_{g0}-T_o(\underline{r}\otimes1_\tau),(\overline{c}-c_0)\otimes 1_\tau,(c_0-\underline{c})\otimes 1_\tau)$
and $\mathcal{C}$, $\mathcal{E}$ are defined in \eqref{matrix}. Here, we use the fact that $\text{P}_{\mathbb{R}^{n\tau}}(\theta)=\theta$.

Based on system \eqref{eal}, we can analyze the convergence of the proposed algorithm. It should be noted that the results derived in this subsection are under the following assumption

\begin{assumption} \label{assup1}
The MTSED problem \eqref{op} satisfies the strong Slater condition \cite{dis4}, i.e., there exists a sufficient small~constant $\varrho>0$ such that $p_{gi}[k]$, $q_{gi}[k]$, $p_{ci}[k]$, $q_{di}[k]$, $i\in\mathcal{N}$,~satisfy~constraints \eqref{c1}, \eqref{c2} and the following conditions for all $k\in\mathcal{T}$
\begin{subequations}
\begin{align}
&\underline p_{gi}+\varrho\leq p_{gi}[k]\leq\overline p_{gi}-\varrho,~\forall i\in\mathcal{N}_g\label{cc3}\\
&{\underline q_{gi}+\varrho\leq q_{gi}[k]\leq \overline q_{gi}-\varrho,~\forall i\in\mathcal{N}_g}\label{cc4}\\
&{\underline r_{i}T_o+\varrho\leq p_{gi}[k]-p_{gi}[k-1]\leq \overline r_{i}T_o-\varrho,~\forall i\in\mathcal{N}_g}\label{cc5}\\
&{\varrho\leq p_{ci}[k]\leq \overline p_{ci}-\varrho,~\forall i\in\mathcal{N}_s}\label{cc6}\\
&{\varrho\leq p_{di}[k]\leq \overline p_{di}-\varrho,~\forall i\in\mathcal{N}_s}\label{cc7}\\
&{\underline c_{i}+\varrho\leq c_{i}[0]+T_o\sum_{l=1}^{k}(\eta_{ci} p_{ci}[l]-\eta_{di}^{-1} p_{di}[l])}\nonumber\\
&\hspace{4.9cm}{\leq\overline c_{i}-\varrho,~\forall i\in\mathcal{N}_s}\label{cc9}\\
&{\underline v_{i}+\varrho\leq v_{i}[k]\leq \overline v_{i}-\varrho,~\forall i\in\mathcal{N}}\label{cc10}
\end{align}
\end{subequations}
\end{assumption}
Now, we give the main results of the paper with respect to convergence of algorithm \eqref{al} in the following theorem.

\begin{theorem} \label{aca}
Given any bounded initial points, the trajectories of system \eqref{al} are bounded and asymptotically converge to an optimal solution of the MTSED problem \eqref{op}.
\end{theorem}

\textit{Proof:}  To simplify the notation, denote $\zeta=\text{col}(x,y,z,\rho)$ and $\zeta^*=\text{col}(x^*,y^*,z^*,\rho^*)$ as the state and equilibrium point of system \eqref{eal}, respectively. Consider the following function
\begin{equation*}\label{vx}
\begin{split}
V(\zeta)=&W(x)+\frac{1}{2}(\Vert y-y^*\Vert^2+\Vert z-z^*\Vert^2+\Vert \rho-\rho^*\Vert^2)\nonumber
\end{split}
\end{equation*}
where $W(x)=\frac{1}{2}\Vert x-\tilde x^*\Vert^2-\frac{1}{2}\Vert x-\tilde{x}\Vert^2$ with $\tilde{x}^*=\text{P}_{\Omega}(x^*)$. Based on Lemma \ref{lemma3}, $W(x)$ has the following properties
\setcounter{equation}{20}
\begin{equation}\label{qt}
\begin{split}
\frac{1}{2}\Vert\tilde{x}-\tilde{x}^* \Vert^2&\leq W\leq\frac{1}{2}\Vert x-\tilde{x}^*\Vert^2,~\frac{\partial W}{\partial x}=\tilde{x}-\tilde{x}^*.
\end{split}
\end{equation}
Therefore, $V\geq 0$. Taking the time derivative of $V$ along the trajectory of system \eqref{eal} gives 
\begin{align}
\dot{V}=&(\tilde x-\tilde{x}^*)^T(-x+\tilde x-\mathcal{A}\tilde x-\mathcal{B}-\mathcal{C}^T(y+\rho)-\mathcal{E}^T{z}^+)\nonumber\\
&+(y-y^*)^T(\mathcal{C}\tilde x-\mathcal{D})+(z-z^*)^T(z^+-z)\nonumber\\
&+(\rho-\rho^*)^T(-\rho+\mathcal{C}\tilde x-\mathcal{D}).\label{dv}
\end{align}
At the equilibrium point $\zeta^*$, system \eqref{eal} satisfies
\begin{subequations}\label{20}
\begin{align}
0&=-x^*+\tilde x^*-\mathcal{A}\tilde x^*-\mathcal{B}-\mathcal{C}^Ty^*-\mathcal{E}^Tz^*\label{eq1}\\
0&=\mathcal{C}\tilde x^*-\mathcal{D}\label{eq3}\\
0&=-z^*+\text{P}_{\mathbb{R}_+^{4n\tau}}(\mathcal{E} \tilde x^*-\mathcal{F}+z^*)\label{eq2}\\
0&=\rho^*.\label{eq4}
\end{align}
\end{subequations}
Then, it holds that
\begin{align}\label{dv1}
\dot{V}=&-(\tilde{x}-\tilde{x}^*)^T(x-\tilde x)+(\tilde{x}-\tilde{x}^*)^T(x^*-\tilde{x}^*)\nonumber\\
&-(\tilde{x}-\tilde{x}^*)^T\mathcal{A}(\tilde{x}-\tilde{x}^*)-(\tilde{x}-\tilde{x}^*)^T\mathcal{E}^T(z^+-z^*)\nonumber\\
&-(z-z^*)^T(z-z^+)-(\rho-\rho^*)^T(\rho-\rho^*).
\end{align}
According to Lemma \ref{lemma4}, by replacing $S$, $\xi$, $\eta$ in \eqref{pr1} with $\Omega$, $x$, $\tilde{x}^*$, respectively, we have
\begin{align}
\label{prxx}
(\tilde{x}-\tilde{x}^*)^T(x-\tilde{x})\geq 0.
\end{align}
Similarly, we can obtain
\begin{subequations}
\begin{align}
(\tilde{x}^*-\tilde{x})^T(x^*-\tilde{x}^*)&\geq 0\label{prx}\\
(z^+-z^*)^T(\mathcal{E}\tilde x-\mathcal{F}+z-z^+)&\geq 0\label{prz}
\end{align}
\end{subequations}
where inequality \eqref{prx} is derived by replacing $S$, $\xi$, $\eta$ in \eqref{pr1} with $\Omega$, $x^*$, $\tilde x$, respectively; and inequality \eqref{prz} is derived by noting that $z^*\in\mathbb{R}_{+}^{4n\tau}$ from \eqref{eq2} as well as Lemma \ref{lemma1}, and replacing $S$, $\xi$, $\eta$ with $\mathbb{R}_+^{4n\tau}$, $\mathcal{E}\tilde x-\mathcal{F}+z$, $z^*$, respectively. Inequality \eqref{prz} can be further rewritten as 
\begin{align}
-(z-z^+)^T(z^+-z^*)\leq (\mathcal{E}\tilde x-\mathcal{F})^T(z^+-z^*).
\end{align}
As a consequence, we have
\begin{equation}
\label{mum}
\begin{split}
&-(\tilde{x}-\tilde{x}^*)^T\mathcal{E}^T(z^+-z^*)-(z-z^*)^T(z-z^+)\\
=&-(\tilde{x}-\tilde{x}^*)^T\mathcal{E}^T(z^+-z^*)-(z^+-z^*)^T(z-z^+)\\
&-\Vert z-z^+\Vert^2\\
\leq&-(\mathcal{E}\tilde{x}-\mathcal{E}\tilde{x}^*)^T(z^+-z^*)+(\mathcal{E}\tilde x-\mathcal{F})^T(z^+-z^*)\\
&-\Vert z-z^+\Vert^2\\
=&(\mathcal{E}\tilde{x}^*-\mathcal{F})^T(z^+-z^*)-\Vert z-z^+\Vert^2.
\end{split}
\end{equation}
Substituting \eqref{prxx}, \eqref{prx} and \eqref{mum} into \eqref{dv1} gives
\begin{align}\label{lyaineq}
\dot{V}\leq&-(\tilde x-\tilde{x}^*)^T\mathcal{A}(\tilde x-\tilde{x}^*)+(\mathcal{E}\tilde{x}^*-\mathcal{F})^T(z^+-z^*)\nonumber\\
&-\Vert z-z^+\Vert^2-\Vert\rho-\rho^*\Vert^2.
\end{align}
In addition, applying Lemma \ref{lemma1} to equation \eqref{eq2} gives
\begin{subequations}
\begin{align}
\mathcal{E}\tilde{x}^*-\mathcal{F}\leq 0\label{28a0}\\
(\mathcal{E}\tilde{x}^*-\mathcal{F})^Tz^*=0.\label{28a}
\end{align}
\end{subequations}
Combining  \eqref{28a0} with $z^+\geq0$ gives 
\begin{equation}\label{28aa}
(\mathcal{E}\tilde{x}^*-\mathcal{F})^Tz^+\leq 0.
\end{equation}
Substituting \eqref{28a} and \eqref{28aa} into \eqref{lyaineq} and noting the fact that matrix $\mathcal{A}$ is positive semi-definite yield
\begin{align}
\dot{V}
%
\leq0.\label{vz}
\end{align}

Given any bounded initial condition $\zeta_0=\zeta(0)$, let $\zeta(t,\zeta_0)$ be the corresponding trajectory of system \eqref{eal}. Define set $\mathcal{M}_{1}$ as $\mathcal{M}_{1}=\{\zeta~|~V(\zeta)\leq V(\zeta_0)\}$ which depends on $\zeta_0$. Further, it has been pointed out in \cite{dis4} that under Assumption 1, the optimal Lagrange~multipliers $\lambda^*_{pi}[k]$, $\lambda^*_{qi}[k]$, $\mu^*_{Mi}[k]$, $\mu^*_{mi}[k]$, $\gamma^*_{Mi}[k]$, $\gamma^*_{mi}[k]$, $\forall i\in$ $\mathcal{N},k\in\mathcal{T}$ of the MTSED problem \eqref{op} are bounded, and thus $y^*$, $z^*$ are bounded. Because $\zeta_0$, $\tilde x^*=\text{P}_\Omega(x^*)$, $y^*$, $z^*$
 and $\rho^*$ are bounded, $V(\zeta_0)$ is  bounded. However, $\mathcal{M}_{1}$ could be unbounded due to  the definition of $W(x)$ in $V(\zeta)$. Therefore, the LaSalle invariant principle that was extensively adopted to prove convergence of distributed convex optimization algorithms in the related literature (e.g., \cite{15,yipeng,cen}) cannot be used here. 
  
To overcome this issue, we will first show that $\zeta(t,\zeta_0)$ is bounded in some compact set, and thus according to Lemma \ref{lemma5}, its positive limit set exists and is nonempty. Then, we will show that  the projection of the $x$ component of any point in the positive limit set is an optimal solution of problem \eqref{op}. As $\zeta(t,\zeta_0)$ converges to its positive limit set, the corresponding projection of the $x$ component in $\zeta(t,\zeta_0)$ converges to the optimal solutions of problem \eqref{op}. 
  
  Firstly, we claim $\zeta(t,\zeta_0)\in\mathcal{M}_{1}$. This is due to the fact $\dot V\leq 0$ which yeilds $V(\zeta(t,\zeta_0))\leq V(\zeta_0)$. Define $\tilde\zeta$ and $\tilde \zeta^*$ as $\tilde\zeta=\text{col}(\tilde x,y,z,\rho)$ and $\tilde \zeta^*=\text{col}(\tilde x^*,y^*,z^*,\rho^*)$. Then, we claim that any  $\tilde\zeta$ in set $\mathcal{M}_{1}$ is bounded. This is because 
\begin{equation*}
\begin{split}
\frac{1}{2}\Vert \tilde{\zeta}-\tilde{\zeta}^*\Vert^2\leq V(\zeta)\leq V(\zeta_0).
\end{split}
\end{equation*}
Moreover, the boundedness of $\tilde{x}$ and $z$ suggests that  $z^+=\text{P}_{\mathbb{R}_+^{4n\tau}}(\mathcal{E}\tilde x-\mathcal{F}+z)$ is bounded. Thus, the term $\tilde x-\mathcal{A}\tilde x-\mathcal{B}-\mathcal{C}^T(y+\rho)-\mathcal{E}^Tz^+$ in the dynamics of $x$ in \eqref{eal} is bounded. Without loss of generality, assume that $\Vert\tilde x-\mathcal{A}\tilde x-\mathcal{B}-\mathcal{C}^T(y+\rho)-\mathcal{E}^Tz^+\Vert\leq m_o$ with some $0<m_o<\infty$. Then,  based on the comparison principle  \cite{ns} and  dynamics of $x$ in \eqref{eal}, we get $\varphi_{1i}(t)\leq x_{i}(t)\leq\varphi_{2i}(t)$, $\forall t\geq 0$, where $x_i(t)$, $i=1,2,\dots,6n\tau$ is the $i$th element of $x(t)$, and $\varphi_{1i}(t)$, $\varphi_{2i}(t)$ are  the solutions of the following ordinary differential equations (ODEs)
\begin{align}\label{cp}
\dot{\varphi}_{1i}=-\varphi_{1i}-m_o,~\dot{\varphi}_{2i}=-\varphi_{2i}+m_o
\end{align}
with the initial conditions $\varphi_{1i}(0)=\varphi_{2i}(0)=x_i(0)$, respectively. Solving the ODEs in \eqref{cp} gives $\varphi_{1i}(t)=x_i(0)e^{-t}-m_o(1-e^{-t})$, and $\varphi_{2i}(t)=x_i(0)e^{-t}+m_o(1-e^{-t})$. Hence, we have
\begin{align}
\Vert x_i(t)\Vert\leq \Vert x_i(0)\Vert+m_o,~\forall t\geq 0
\end{align}
which implies that the component  $x(t)$ in $\zeta(t,\zeta_0)$ is bounded. Therefore, $\zeta(t,\zeta_0)$ belongs to $\mathcal{M}_{1}$ and  is bounded.  

Without loss of generality, suppose $\Vert\zeta(t,\zeta_0)\Vert\leq M_0$ with some $0<M_0<\infty$. Since $\tilde {\zeta}^*$ is also bounded, we can define $\mathcal{M}_{2}$ as $\mathcal{M}_{2}=\{\zeta~|~\Vert\zeta-\tilde{\zeta}^*\Vert\leq M\}$ with some $0<M<\infty$. So, $\mathcal{M}_{2}$ is a compact set, and $\zeta(t,\zeta_0)\in\mathcal{M}_{2}$.  Then, according to Lemma \ref{lemma5}, the positive limit set $\mathcal{L}_{\zeta_0}^+$ of $\zeta(t,\zeta_0)$ is nonempty, compact and invariant; and it is  in $\mathcal{M}_{2}$ as $\mathcal{M}_{2}$ is a closed set. Moreover,  $\zeta(t,\zeta_0)$ approaches $\mathcal{L}_{\zeta_0}^+$ as $t\to+\infty$.
%

Now, we will show that the projection of the $x$ component of any point in $\mathcal{L}_{\zeta_0}^+$  is an optimal solution of problem \eqref{op}. Since $V(\zeta(t,\zeta_0))$ is a decreasing function of $t$ ($\dot{V}\leq0$) and is continuous on the compact set $\mathcal{M}_{2}$, it is bounded from below on $\mathcal{M}_{2}$. Therefore, $V(\zeta(t,\zeta_0))$ has a limit $\kappa$ as $t\to+\infty$.  By noting that $\mathcal{L}_{\zeta_0}^+\subset\mathcal{M}_{2}$, for any $\chi\in\mathcal{L}_{\zeta_0}^+$, there is a sequence $t_m$ with $m\to+\infty$ such that $\lim_{m\to\infty}\zeta(t_m,\zeta_0)\to \chi$. Due to the continuity of $V(\zeta(t,\zeta_0))$, $V(\chi)=\lim_{m\to\infty}V(\zeta(t_m,\zeta_0))=\kappa$. Hence, $V(\zeta)=\kappa$ on $\mathcal{L}_{\zeta_0}^+$. By recalling the fact that  $\mathcal{L}_{\zeta_0}^+$ is an invariant set, $\dot{V}(\zeta)=0$ on  $\mathcal{L}_{\zeta_0}^+$. Therefore, we have 
\begin{equation}
\mathcal{L}_{\zeta_0}^+\subset\mathcal{M}_3\subset\mathcal{M}_2
\end{equation}
with $\mathcal{M}_3=\{\zeta\in\mathcal{M}_2~|~\dot{V}(\zeta)=0\}$.

Let $U(\zeta)=-(\tilde x-\tilde{x}^*)^T\mathcal{A}(\tilde x-\tilde{x}^*)+(\mathcal{E}\tilde{x}^*-\mathcal{F})^T(z^+-z^*)-\Vert z-z^+\Vert^2-\Vert\rho-\rho^*\Vert^2$, and define $\mathcal{M}_4$ as follows
\begin{align}
\mathcal{M}_4=\{\zeta~|~U(\zeta)=0\}
\end{align}
From \eqref{lyaineq}, we have $\dot V(\zeta)\leq U(\zeta)$, and thus 
\begin{equation}
\mathcal{M}_3\subseteq\mathcal{M}_4.\label{343}
\end{equation}
Based on \eqref{28a} and \eqref{28aa}, $U(\zeta)=0$ is equivalent to 
\begin{subequations}\label{27}
\begin{align}
&(\tilde{x}-\tilde{x}^*)^T\mathcal{A}(\tilde{x}-\tilde{x}^*)=0\\
&(\mathcal{E}\tilde{x}^*-\mathcal{F})^T(z^+-z^*)=0\\
&z^+-z=0\label{34c}\\
&\rho=\rho^*=0.\label{34b}
\end{align}
\end{subequations}
From \eqref{34c} and Lemma \ref{lemma1}, we have
\begin{align}\label{lianga}
{z}^T(\mathcal{E}{\tilde{x}}-\mathcal{F})=0
\end{align}
{Since $\rho=\rho^*\equiv 0$ on $\mathcal{M}_4$, for any $\zeta \in \mathcal{L}_{\zeta_0}^+$, we have $\dot{\rho}=0$ by noting the facts that $\mathcal{L}_{\zeta_0}^+\subset\mathcal{M}_4$ from \eqref{343} and $\mathcal{L}_{\zeta_0}^+$ is an invariant set}. Then, $\dot\rho=0$ together with  \eqref{eal} gives
\begin{align}\label{liangb}
\mathcal{C}{\tilde{x}}-\mathcal{D}=0
\end{align}
 Due to \eqref{28a}, we have $(\mathcal{E}\tilde{x}^*-\mathcal{F})^Tz=0$, which together with \eqref{lianga} gives
\begin{align}
({\tilde{x}}-\tilde{x}^*)^T\mathcal{E}^T{z}=0\label{35}.
\end{align}
From \eqref{dv}, \eqref{20}, \eqref{27}, \eqref{lianga}, \eqref{liangb}  and \eqref{35}, we have
\begin{align}
\dot{V}(\zeta)=&(\tilde{x}-\tilde{x}^*)^T(-x+\tilde{x}-\mathcal{A}\tilde{x}-\mathcal{B})-(\mathcal{C}\tilde{x}-\mathcal{C}\tilde{x}^*)^T(y+\rho)\nonumber\\
&-(\tilde{x}-\tilde{x}^*)^T\mathcal{E}^Tz^++(y-y^*)^T(\mathcal{C}\tilde x-\mathcal{D})\nonumber\\
&+(z-z^*)^T(z^+-z)+(\rho-\rho^*)^T(-\rho+\mathcal{C}\tilde x-\mathcal{D})\nonumber\\
=&(\tilde{x}-\tilde{x}^*)^T(-x+\tilde{x}-\mathcal{A}\tilde{x}-\mathcal{B})
\end{align}
for all $\zeta\in\mathcal{M}_4$.
Furthermore, for all $\zeta\in\mathcal{M}_3\subseteq\mathcal{M}_4$, we have $\dot{V}(\zeta)=0$, i.e., 
\begin{align}\label{36}
(\tilde{x}-\tilde{x}^*)^T(-x+\tilde{x}-\mathcal{A}\tilde{x}-\mathcal{B})=0
\end{align}
which is equivalent to
\begin{align}\label{37}
(\tilde{x}-\tilde{x}^*)^T(\mathcal{A}\tilde {x}+\mathcal{B})=-({\tilde{x}}-\tilde{x}^*)^T(x-{\tilde x})\leq 0.
\end{align}
The inequality in \eqref{37} is derived from \eqref{prxx}. Denote $f(\tilde{x})=\sum\nolimits_{i=1}^n\sum\nolimits_{k=1}^{\tau}(f_{gi}(\tilde{p}_{gik})+f_{si}(\tilde{p}_{cik},\tilde{p}_{dik}))$ as the total cost function. For any point $\zeta\in\mathcal{M}_3$, let $\hat{\tilde x}$ be the projection of the $x$ component of $\zeta$. Since $f(\tilde x)$ is convex on $\tilde x$,  we have $f(\hat{\tilde{x}})-f(\tilde{x}^*)\leq(\nabla f(\tilde{x})|_{\tilde{x}=\hat{\tilde{x}}})^T(\hat{\tilde{x}}-\tilde{x}^*)$. Further, since $\nabla f(\tilde x)|_{\tilde x=\hat{\tilde{x}}}=\mathcal{A}\hat{\tilde{x}}+\mathcal{B}$, we have $f(\hat{\tilde{x}})-f(\tilde{x}^*)\leq(\mathcal{A}\hat{\tilde{x}}+\mathcal{B})^T(\hat{\tilde{x}}-{\tilde{x}}^*)\leq 0$ which gives rise to $f(\hat{\tilde{x}})\leq f(\tilde{x}^*)$. On the other hand, $\tilde{x}^*$ is an optimal solution of \eqref{op}, hence, $f(\tilde{x}^*)\leq f(\hat{\tilde{x}})$. Then, we have $f(\hat{\tilde{x}})=f(\tilde{x}^*)$, which implies $\hat {\tilde{x}}$ is also an optimal solution to problem \eqref{op}. Due to the arbitrariness of $\hat{\tilde{x}}$, i.e., $\forall \zeta\in\mathcal{M}_3$, we conclude that the projection $\tilde{x}$ of the component of any $\zeta\in\mathcal{M}_3$ is an optimal solution to problem \eqref{op}. Since $\mathcal{L}_{\zeta_0}^+\subset\mathcal{M}_3$, the projection of the $x$ component of any point in $\mathcal{L}_{\zeta_0}^+$  is an optimal solution of problem \eqref{op}. Moreover,  $\zeta(t,\zeta_0)$ approaches $\mathcal{L}_{\zeta_0}^+$ as $t\to\infty$.  This completes the proof. $\hfill\blacksquare$

 
\begin{remark}
The LaSalle invariance principle usually plays a crucial role in convergence analysis in the literature of distributed convex optimization (e.g., \cite{15,yipeng,cen,sf,dis4}). However, it cannot be used in the proof of Theorem \ref{aca}. This is due to the fact that the positive invariant set $\mathcal{M}_1$ might be unbounded, which results from the particular form of $V(\zeta)$.~To address this issue, we use another property of system  \eqref{al}, i.e., its trajectory with any bounded initial condition is bounded. Then, we  apply properties of positive limit sets   (Lemma \ref{lem5}) and a similar idea in the  proof of the LaSalle invariance principle \cite{ns} to get the convergence  of \eqref{al}. 
\end{remark}

\begin{remark} ED is usually performed periodically based on the latest net demand forecast in practice, but Theorem \ref{aca} gives the asymptotic convergence of the proposed algorithm.  Hence,  to implement our MTSED scheme, a criterion for terminating algorithm \eqref{al} in finite time is needed. A typical stopping rule, i.e., set a predefined maximum allowable computational time (e.g., 3 mins), has been extensively used in ED (e.g., \cite{oc1,tmc,tmc1,tmc2}), and thus is adopted in this paper as well. Here, it should be noted that the control performance of~MTSED~could~deteriorate if \eqref{al} does not converge in some cases.  However,~as~we show in Section IV,  the distributed~algorithm~is~able~to~converge to an optimal solution of \eqref{op} within the above maximum computational time, and thus is well performed in achieving fast convergence (at least) for systems considered in the case study. Of course, how to build an algorithm that has a theoretic guarantee on the convergence rate, i.e., ensure a satisfactory apriori bound on convergence time,  still remains open and should be studied in the future.
\end{remark}


\begin{remark}\label{market}
Many countries have deregulated electricity markets, e.g., the EU Electricity Market \cite{market1}, which effectively transform the MTSED from cost-based operations with a goal of cost minimization to bid-based operations with a goal of social welfare maximization where the users' flexible effects are taken into account.  It is worth pointing out that the derived results still apply to the bid-based problem. On the other hand, the forecast uncertainty of renewable generation and load demand is another key issue that should be considered in power system operation. To incorporate forecast uncertainties with MTSED, a typical approach that is widely used in the literature is chance-constrained programming, where constraints can be violated with a relatively small level of probability. In fact, the bid-based chance-constrained MTSED problem in deregulated electricity markets which takes stochastic renewable output and demand predictions into account, can also be formulated by \eqref{op}  with some minor modifications as follows (see \cite{market1,market2} for more details of the bid-based MTSED)
\begin{subequations}\label{bid}
\begin{align}
\mathop{\textup{max}}~&\sum_{i=1}^n\sum_{k=1}^{\tau}(f_{l_i}(d_{p_i}[k])-f_{gi}(p_{gi}[k])-f_{si}(p_{ci}[k],p_{di}[k]))\nonumber\\
\text{s.t.}~&\text{Constraints \eqref{c4}, \eqref{c5}, \eqref{c9} and \eqref{c10}}\nonumber\\
&P_r\Big\{p_{gi}[k]-p_{ci}[k]+p_{di}[k]-\sum_{j=1}^n (g_{ij}v_{j}[k]-b_{ij}'\theta_{j}[k])\nonumber\\
&\geq d_{pi}[k]\Big\}\geq P_{rp},~\forall i\in\mathcal{N},k\in\mathcal{T}\label{bid1}\\
&P_r\Big\{q_{gi}[k]+\sum_{j=1}^n(b_{ij}v_{j}[k]+g_{ij}\theta_{j}[k])\geq d_{qi}[k]\Big\}\nonumber\\
&\geq P_{rq},~\forall i\in\mathcal{N},k\in\mathcal{T}\label{bid2}\\
&\underline p_{gi}[k]\leq p_{gi}[k]\leq\overline p_{gi}[k],~\forall i\in\mathcal{N}_g,k\in\mathcal{T}\\
&\underline d_{pi}[k]\leq d_{pi}[k]\leq \overline d_{pi}[k],~\forall i\in\mathcal{N},k\in\mathcal{T}\\
&0\leq p_{ci}[k]\leq \overline p_{ci}[k],~\forall i\in\mathcal{N}_s,k\in\mathcal{T}\\
&0\leq p_{di}[k]\leq \overline p_{di}[k],~\forall i\in\mathcal{N}_s,k\in\mathcal{T}
\end{align}
\end{subequations}
where $\underline p_{gi}[k]$, $\overline p_{gi}[k]$, $\underline d_{pi}[k]$, $\overline d_{pi}[k]$, $\overline p_{ci}[k]$, $\overline p_{di}[k]$ are defined similar as those in problem \eqref{op} but represent the respective bid limits;$P_{r}\{\cdot\}$ denotes the probability of the argument to hold. Functions $f_{gi}(p_{gi}[k])$, $i\in\mathcal{N}_g$, $f_{si}(p_{ci}[k],p_{di}[k])$, $i\in\mathcal{N}_s$, and $f_{l_i}(d_{p_i}[k])$, $i\in\mathcal{N}$ are bid functions of each SG, load and ESD in the electricity market, where $f_{gi}(p_{gi}[k])$, $f_{si}(p_{ci}[k],p_{di}[k])$ are defined in \eqref{gc}, \eqref{esc} (i.e., the same as the MTSED problem \eqref{op}), and 
\begin{equation}
f_{li}(d_{pi}[k])=\frac{a_{di}}{2}  d^2_{pi}[k]+b_{di} p_{gi}[k]+c_{di}
\end{equation}
with $a_{di}<0$, $b_{di},c_{di}\in\mathbb{R}$, $\forall i\in\mathcal{N}$. In \eqref{bid}, chance constraints \eqref{bid1}, \eqref{bid2} indicate that the total active (reactive) power increment of each bus need to meet the local net demand at each time slot with a predefined confidence level of $P_{rp}$ ($P_{r_q}$). As argued in \cite{sf}, various probability distribution functions (e.g., normal distribution function) can be used to account for the renewable generation and load forecast errors. Under these certain distribution functions, the chance constraints \eqref{bid1}, \eqref{bid2} can be converted into deterministic equivalent linear inequalities. Therefore,  problem \eqref{bid} is still a convex problem with linear feasibility constraints, and thus the proposed method applies.
\end{remark}

\section{Case Study}
In this section, we test the effectiveness of the proposed distributed algorithm on a modified IEEE 14-bus system in which we add ESDs at buses 2, 5, 7,~9, 10, 12 and 13. The diagram of the test system is shown in Fig. 1.  For simplicity, we assume that each ESD is subject to the same linear cost coefficients $a_{si}=10.5$ \$/MWh, $b_{si}=120$ \$/h, power limits $\overline{p}_{ci}=\overline{p}_{di}=25$ MW, and energy capacity limits $\underline{c}_i=1.25$ MWh, $\overline{c}_{i}=25$ MWh. Moreover, we set $c_{i}[0]=6.25$ MWh, $\eta_{ci}=0.95$, and $\eta_{di}=0.9$ for all ESDs. 

\begin{figure}[t]
\centering
\epsfig{figure={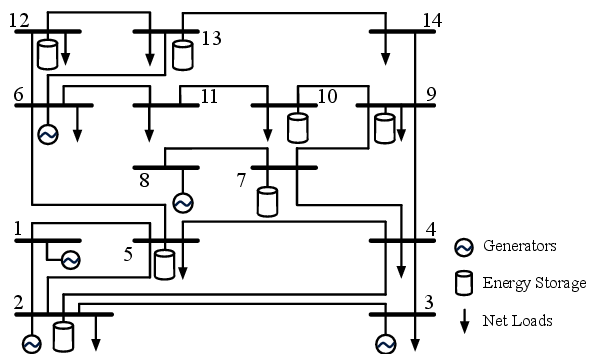},width=0.8\linewidth}
\caption{Diagram for the modified IEEE 14-bus power system.}
\end{figure}
\begin{table}[t]
\caption{Generator Parameters}
\begin{center}\tiny
\setlength{\tabcolsep}{2.5pt}
\renewcommand\arraystretch{1.6}
\begin{tabular}{c|cccccccccccc}
\hline
\hline
Bus&$a_{gi}$&$b_{gi}$&$c_{gi}$&$\overline{p}_{gi}$&$\underline{p}_{gi}$&$\overline{q}_{gi}$&$\underline{q}_{gi}$&$\overline{r}_i$&$\underline{r}_i$\\
&$\$/(\text{MW}^2\cdot\text{h})$&$\$/\text{MWh}$&$\$$&MW&MW&MW&MW&MW/h&MW/h\\
\hline
1&0.014&7&240&332&0&10&0&250&80\\
\hline
2&0.019&10&200&140&0&50&0
&120&50\\
\hline
3&0.018&8.5&220&100&0&40&0&150&65\\
\hline
6&0.018&11&200&100&0&24&0&150&50\\
\hline
8&0.016&10.5&220&100&0&24&0&120&50\\
\hline
\hline
\end{tabular}
\end{center}
\end{table}

\begin{figure}[t]
\centering
\subfigure[The time evolution of $\tilde{\lambda}_{pk1}$ (p.u.).]
{\includegraphics[width=0.8\linewidth]{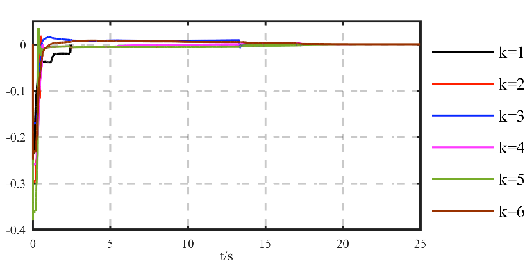}}
\subfigure[The time evolution of $\tilde{\lambda}_{qk1}$ (p.u.).]
{\epsfig{figure={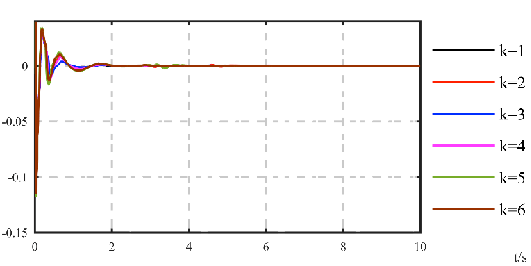},width=0.8\linewidth}}
\caption{State responses of the proposed algorithm}
\end{figure}
 
The prediction window for the MTSED problem is set to be $1$ h which is divided into 6 time slots, i.e.,  $\tau=6$ and $T_o=10$ mins.  For bus $i$, $i=2,3,4,5,6,9,10,11,12,13,14$, we set its active and reactive powers of the net demand for the whole prediction window as $(21.7+i,29.5+i,13.5+i,14.9+i,37.8+i,21.3+i)$ MW, and $(2.2+0.5i,1.8+0.5i,0.8+0.5i,1.5+0.5i,2.5+0.5i,2.3+0.5i)$ MVar, respectively. Further, the cost function coefficients, power capacity limits, and ramping limits of all SGs are given in Table \uppercase\expandafter{\romannumeral1}, which are adopted from \cite{dis4} with some modifications.  The power output $p_{gi}[0]$ of each SG is set equally as $p_{gi}[0]=50$ MW. Moreover, we assume the voltage magnitude bounds are the same for all buses, i.e., $\underline{v}_{i}=0.9$ p.u. and $\overline{v}_{i}=1.1$ p.u..

We take the time evolution of the auxiliary variables $\tilde{{\lambda}}_{pi}$, $\tilde{{\lambda}}_{qi}$ of bus $1$ for an instance to show the asymptotic convergence of the proposed algorithm. It can be observed from Fig. 2 that $\tilde{{\lambda}}_{p1}$, $\tilde{{\lambda}}_{q1}$ converge to $0$ asymptotically. This, together with \eqref{al7}, \eqref{con1a} and \eqref{con1b} indicates that the active and reactive power balance constraints \eqref{c1}, \eqref{c2} hold for all time slots at the steady state of the designed dynamical system. 
   
The detailed converged values of the proposed algorithm with respect to the active/reactive power outputs of all synchronous generating units, charging/discharging powers of all ESDs and voltage magnitudes of all buses across the whole prediction window are given Fig. 3.  It shows in the figure that only SGs contribute to the system power balance at times slots 1,  4 and 6. This is because SGs usually have smaller marginal costs compared with storage and are preferred from a cost-effectiveness viewpoint.  ESDs are charged at time slot 3,  and discharged at time slots 2, 5, which is due to the fact that the corresponding total net demand fluctuates dramatically and SGs do not have enough ramping capacities. Further,  it shows in the figure that the power capacity limits \eqref{c3}, \eqref{c4}, \eqref{c6}, \eqref{c7} of all controllable units and bus voltage amplitude limit \eqref{c10} are satisfied for all time slots.  In particular,  through simulations, we find that the voltages of buses 12 and 14 are close to their lower bounds in Fig.  3,  but out of the acceptable ranges while removing voltage constraints from MTSED (the corresponding simulation results are omitted due to the space issue), which shows the effectiveness of our algorithm in voltage regulation.

In addition, it is shown in Fig. 4 that the converged values fulfil the generator ramping constraint \eqref{c5} as well as storage energy constraint \eqref{c9}. Thus, the converged solution of the proposed algorithm satisfies all feasibility conditions of the MTSED problem \eqref{op}. Moreover, during the simulation period, the total operation cost converges to a minimum 32642.9 \$/h over the whole prediction window, which coincides with the results derived by the well-known CVX solver in a centralized way \cite{sc}. Therefore, the developed distributed projection-based algorithm converges to an optimal solution of the MTSED problem \eqref{op}.

\begin{figure}[t]
\centering
\subfigure[The optimal active power (left) and reactive power (right) of SGs (in unit of MW).]{\includegraphics[width=0.8\linewidth]{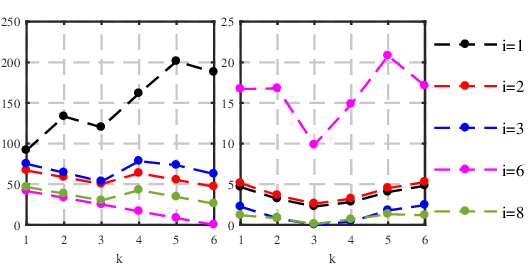}}
\subfigure[The optimal charging power (left) and discharging power (right) of ESDs (in unit of MW).]{\epsfig{figure={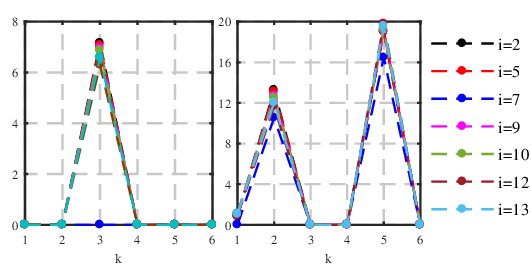},width=0.8\linewidth}}
\subfigure[Bus voltage magnitudes (p.u.).]
{\epsfig{figure={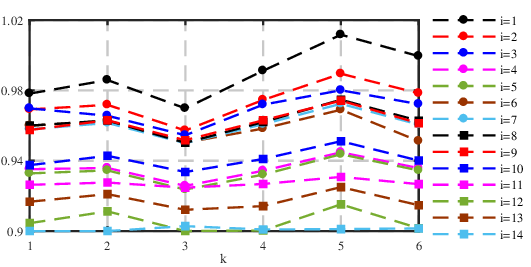},width=0.8\linewidth}}
\caption{The converged values of the proposed algorithm across all time slots in the prediction window.}
\end{figure}

\begin{figure}[h]
\centering
\subfigure[The ramping rates (in unit of MW/h) of SGs.]
{\includegraphics[width=0.8\linewidth]{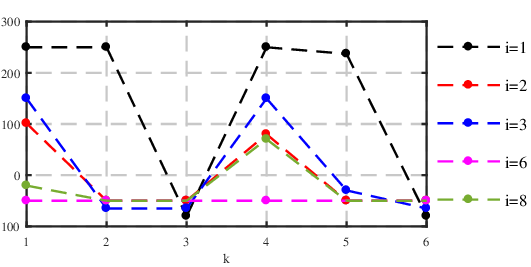}}
\subfigure[The energy levels (in unit of MWh) of ESDs.]{\epsfig{figure={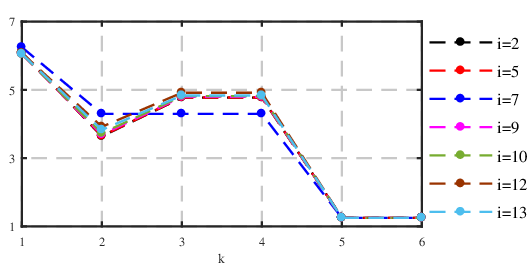},width=0.8\linewidth}}
\caption{The ramping rates of SGs and energy stored in ESDs across all times slots in the prediction window}
\end{figure}

%
%
\section{Conclusions}
This paper has studied the multi-time slot economic dispatch problem of power networks that aims to optimally coordinate the active/reactive powers produced by SGs and charging/discharging powers of ESDs to meet a net demand profile over a receding finite time horizon while respecting the system operational constraints. To solve the problem in a distributed way, a  projection-based algorithm depending on information that each bus can obtain~has~been proposed. Simulation studies have been conducted on a modified IEEE 14-bus system, which has validated the effectiveness of the proposed method.


\bibliographystyle{plain}        

\end{document}